# Study of magneto-thermal resistance effect in a Co$_{50}$Fe$_{50}$/Cu multilayer through the analysis of electron and lattice thermal conductivities

Fuya Makino,[1,2,3] Takamasa Hirai,[2,*] Takuma Shiga,[4] Hirofumi Suto,[2] Hiroshi Fujihisa,[4] Koichi Oyanagi,[3] Satoru Kobayashi,[3] Taisuke Sasaki,[3] Takashi Yagi,[4] Ken-ichi Uchida,[1,2,5] and Yuya Sakuraba[1,2,†]

[1]*Graduate School of Science and Technology, University of Tsukuba, Tsukuba, Ibaraki, Japan.*

[2]*National Institute for Materials Science (NIMS), Tsukuba, Ibaraki, Japan.*

[3] *Faculty of Science and Engineering, Iwate University, Morioka, Iwate, Japan.*

[4] *National Institute of Advanced Industrial Science and Technology (AIST), Tsukuba, Ibaraki, Japan.*

[5] *Department of Advanced Materials Science, Graduate School of Frontier Sciences, The University of Tokyo, Kashiwa, Chiba, Japan.*

[*]HIRAI.Takamasa@nims.go.jp, [†]SAKURABA.Yuya@nims.go.jp

**ABSTRACT**. This study investigates the giant magneto-thermal resistance (GMTR) effect in a fully-bcc epitaxial Co$_{50}$Fe$_{50}$/Cu multilayer through both experimental and theoretical approaches. The applied magnetic field results in a giant change of the cross-plane thermal conductivity ($\Delta\kappa$) of 37 W m$^{-1}$ K$^{-1}$, which reaches 1.5 times larger than the previously reported value for a magnetic multilayer and record the highest value at room temperature among the other solid-state thermal switching materials working on different principles. We investigated the electron thermal conductivity for exploring the remarkable $\Delta\kappa$ by the two-current-series-resistor model combined with the Wiedemann-Franz (WF) law. However, the result shows the electron contribution accounts for only 35% of the $\Delta\kappa$, indicating the presence of additional spin-dependent heat carriers. Further investigation of the lattice thermal conductivity, which is expected to be spin-independent, using non-equilibrium molecular dynamics (NEMD) simulations suggests a striking contrast: the additional spin-dependent heat carrier contribution is significantly enhanced in the parallel magnetization configuration but nearly negligible in the antiparallel configuration. These findings provide a fundamental insight into the origin of large GMTR effect and highlight its potential of active thermal management technologies for future electronic devices.

## I. INTRODUCTION

Thermal switching devices are pivotal for advancing thermal management technologies, which are essential for enhancing the efficiencies in modern electronic devices [1,2]. By controlling heat flow through a dynamic change in thermal conductivity of solids, these devices address critical challenges in managing heat dissipation across nanoscale to macroscale regimes. To realize practical thermal switching, it is necessary to achieve a significant change of thermal conductivity, a wide operating temperature range, and seamlessly integration with existing electronic systems. Various mechanisms yielding the thermal conductivity change in solids have been extensively investigated so far, including metal-insulator transition [3,4], electrochemical reaction [5,6], voltage-controlled ferroelectricity [7,8], superconductors [9,10] and the magnetoresistance (MR) effect [11]. Among them, the MR effect in magnetic materials offers distinct advantages; it enables noncontact, high-durability and high-scalable thermal conductivity switching by applying an external magnetic field and/or controlling a magnetization of magnetic materials. However, the magnetic-filed-induced change for the thermal conductivity in single magnetic materials is much smaller than other mechanisms at room temperature, hindering the practical application of MR-based thermal switching. [11].

In magnetic multilayers consisting of alternatively stacked ferromagnetic metal (FM) and non-magnetic metal (NM) layers, their electrical conductivity depends on the relative magnetization angle of the adjacent ferromagnetic layers, a phenomenon known as the giant magnetoresistance (GMR) effect [12,13]. The GMR effect is one of the most familiar phenomena in spintronics and has been widely investigated for various applications, such as read heads for hard disk drives and magnetic sensors. Similar to electrical conductivity, thermal conductivity is influenced by the magnetization configuration in these multilayers, leading to a phenomenon called the giant magneto-thermal resistance (GMTR) effect [14-16]. The GMTR effect enables a large change of the thermal conductivity from the parallel (P) and antiparallel (AP) magnetization configurations. The performances of the GMR and GMTR effects are commonly characterized by the MR ratio and magneto-thermal resistance (MTR) ratio, defined as $(\sigma_{AP} - \sigma_P)/\sigma_{AP}$ and $(\kappa_P - \kappa_{AP})/\kappa_{AP}$, respectively. Here, $\sigma_{AP(P)}$ and $\kappa_{AP(P)}$ represent the electrical conductivity $\sigma$ and the thermal conductivity $\kappa$ in the AP state (P state), respectively. It is known that $\sigma$ and $\kappa$ in the metallic materials and systems are often well connected through Wiedemann-Franz (WF) law, i.e., the linear relationship between the electron thermal conductivity and the electrical conductivity. Additionally, thermal conductivity has lattice contribution. Therefore, when only these two contributions are assumed, the MTR ratio should be smaller than the MR ratio because the lattice component should be independent of magnetization direction. Previous studies also suggested that in the presence of inter-spin and spin-conserving inelastic scattering in metallic magnetic multilayers where the whole thermal transport is explained by the contribution of electrons, the WF law does not hold and the MTR ratio can be further reduced [17,18]. However, Nakayama et al. recently reported that the MTR ratio (~150%) is significantly higher than the MR ratio (~60%) in a fully-bcc epitaxial $Co_{50}Fe_{50}$ (3 nm)/Cu (1.6 nm) multilayer, in which Cu forms a metastable bcc structure at room temperature [19,20]. To explain this behavior, they investigated the applicability of the WF law in the case where the density of state shows a steep change near the Fermi energy [21] and performed the first-principles calculation of the spin-dependent ballistic electron transmittance to analyze the GMTR effect. Although their calculated MTR ratio can be larger than MR ratio at the certain temperature range, this range does not include room temperature and the difference between MTR and MR ratios was too small to fully explain the measured MTR ratio. The experimentally observed large MTR ratio holds promise for future thermal management applications and raises fundamental interest in uncovering additional contribution of thermal transport. In this respect, the direct comparison between MTR and MR ratios in the cross-plane direction is essential to understand the electrical contribution. However, and such comparison have so far been limited to the fully-bcc epitaxial $Co_{50}Fe_{50}$ (3 nm)/bcc-Cu (1.6 nm) multilayer system. Furthermore, the quantitative analysis of the contribution of the lattice thermal conductivity arising from phonons, one of essential heat carriers, remains unexplored in such a magnetic multilayer.

In this study, we observed the GMTR effect in a $Co_{50}Fe_{50}$/metastable bcc-Cu multilayer with a thicker $Co_{50}Fe_{50}$ layer (5.1 nm) and compared the behavior to that with a thinner $Co_{50}Fe_{50}$ layer (3.0 nm) used in the previous study [19]. Notably, the change in the cross-plane thermal conductivity of the present sample $\Delta\kappa = \kappa_P - \kappa_{AP}$ and MTR ratio were 37 W m$^{-1}$ K$^{-1}$ and 108% respectively; the obtained $\Delta\kappa$ exhibits the highest value among GMTR reported so far. We compared the observed MTR ratio with the MR ratio evaluated by the two-current-series-resistor (2CSR) model [22] in the cross-plane direction to gain insight into the electrical contribution. In addition, we analyzed the atomic resolution micro-structure of $Co_{50}Fe_{50}$ and bcc-Cu using high-angle annular dark field scanning transmission electron microscopy (HAADF-STEM) and calculated the phonon contribution using a nonequilibrium molecular dynamics (NEMD) simulation [23,24] based on the observed microstructure. Our quantitative analyses revealed that approximately 42% of the $\kappa_P$ value and 65% of the $\Delta\kappa$ value were arising from other spin-dependent contributions that are neither the contributions of pure electron nor phonon transports based on the conventional theories.

**II. METHODS**

The magnetic multilayer film with the structure of $Co_{50}Fe_{50}/[bcc\text{-}Cu/Co_{50}Fe_{50}]_{20}$ was deposited at room temperature onto a [001]-oriented MgO single crystalline substrate using an automated ultra-high-vacuum magnetron sputtering system (hereafter, $Co_{50}Fe_{50}$ and bcc-Cu are referred to as CoFe and Cu for simplicity.) The CoFe thickness was designed to be thicker that in the previous study [19], and the Cu thickness was fixed at 2.0 nm to achieve an AP magnetization configuration at zero field via the anti-ferromagnetic interlayer exchange coupling (IEC) between the CoFe layers through the Cu layers. The surface of the multilayer was covered by an Al layer without breaking the chamber vacuum. The Al layer serves as a transducer for the thermal conductivity measurement using the time-domain thermoreflectance (TDTR) method. The actual thickness values of each layer will be shown later. Prior to the deposition of the bottom CoFe layer, the MgO substrate was heated up to 600°C in the sputtering chamber for the surface cleaning. After the deposition of all layers, the sample was annealed at 250°C with applying a constant magnetic field of 300 mT for 1 h [20].

The crystal structure of the multilayer film was characterized by the x-ray diffraction (XRD) with the Cu K$_\alpha$ radiation and HAADF-STEM with the acceleration voltage of 200 kV. The GMR effect in the current-in-plane configuration and the magnetization were measured using a standard dc four probe method and a vibrating sample magnetometer, respectively, by applying a magnetic field parallel to the in-plane direction. To investigate the GMTR effect in the CoFe/[Cu/CoFe]$_{20}$ multilayer film, we performed the TDTR measurement [25,26], one of the optical pump-probe methods, in the front-heating and front-detection configuration. In this method, ultrafast pump laser pulses heat the surface of the Al transducer layer on the CoFe/[Cu/CoFe]$_{20}$ multilayer film, while probe laser pulses irradiated with controlled delay time detect the transient response of the surface temperature via thermoreflectance, i.e., the temperature dependence of the reflectivity, which enables the quantitative determination of thermal transport properties of thin films. Further details on the TDTR experiment are found in Refs [27-29]. All the TDTR measurements were performed at ambient temperature under air atmosphere.

**III. RESULTS**

An out-of-plane XRD pattern of the sample is shown in Fig. 1(a). The diffraction peak with the fringe around 66° corresponds to the 002 peaks of bcc-CoFe, confirming the [001]-oriented growth of bcc-CoFe on the MgO substrate and the formation of the smooth interfaces with a small roughness. Since the peaks from a normal fcc-phase of Cu were not observed, Cu layer was expected to grow in bcc-structure with the [001]-orientated growth in between CoFe layers in our multilayer. Because the lattice constants of bcc-CoFe and bcc-Cu are almost the same, the 002 diffraction peaks originated from bcc-CoFe and Cu cannot be distinguished. Moreover, the 011 peak from the Al layer is overlapped with the 002 peak from the MgO substrate. Figure 1(b) shows a schematic of the fabricated multilayer film with the actual thicknesses confirmed by the microstructure analysis using a transmission electron microscopy (TEM): 5.1 nm for CoFe layers, 2.0 nm for Cu layers, and 44 nm for the Al layer. We confirmed each CoFe layer was thicker than that in the previous study (3.0 nm). The TEM analysis also confirmed that the total thickness $t_{\text{total}}$ of the CoFe/[Cu/CoFe]$_{20}$ multilayer was 147 nm. Figures 1(c)-1(f) show the cross-sectional HAADF-STEM images captured along the [100] zone axis of the MgO substrate. Figure 1(c) indicates that the 20-period CoFe/Cu multilayer structure maintained the interfacial flatness consistently from the bottom to the top layers. To investigate the interfacial atomic lattice matching at different regions [upper, middle, and lower regions in Fig. 1(c)] of the CoFe/[Cu/CoFe]$_{20}$ multilayer, high-magnification HAADF-STEM images and inverse fast Fourier transform images reconstructed from 110 peaks are separately shown in Figs. 1(d)-(f). We directly observed the metastable bcc-Cu that formed in the multilayer from the HAADF-STEM images. Besides, very smooth CoFe/Cu interfaces was observed across all regions. Therefore, we concluded that the nearly complete lattice matching and the atomically flat interfaces were formed in the CoFe/[Cu/CoFe]$_{20}$ multilayer throughout the entire structure. These microstructure analyses validate that our multilayer can be regarded as the homogenous single medium to evaluate its effective thermal conductivity by the TDTR measurements and analyses.

Figures 2(a) and 2(b) respectively show the external magnetic field $H$ dependence of the current-in-plane MR ratio and the normalized magnetization $M/M_s$, where $M_s$ represents the saturation magnetization. Clear resistance plateaus observed in the range of $|\mu_0 H|$ from 2 mT to 15 mT indicate that the AP state was realized by IEC within this magnetic field range, where $\mu_0$ is the vacuum permeability.. At $|\mu_0 H|$ increases beyond 15 mT, the resistance notably drops, and both the MR ratio and magnetization showed the saturation behaviors, indicating the formation of the P state. It is worth noting that the gradual changes in the magnetizations in the field range corresponding to the plateau of the MR curves are likely caused by the formation of magnetic domains which affect the magnetization but not MR ratio, as the AP state is locally formed in each magnetic domain.

TDTR signals were recorded as a function of delay time between pump and probe laser pulses. As shown in Fig. 2(c), the TDTR signal was significantly changed from the AP state ($\mu_0 H = 0$ mT) to the P state ($\mu_0 H = \pm 75$ mT), suggesting the large magnetization-configuration-dependent thermal conduction. By analyzing TDTR signals using the one-dimensional heat diffusion model, we experimentally estimated the effective thermal conductivity of the CoFe/[Cu/CoFe]$_{20}$ multilayer ($\kappa_{\text{exp}}$) (see Supplementary Material [27]). The fitting curves [solid curves in Fig. 2(c)] showed good agreement with the experimental results. Figure 2(e) summarizes the $H$ dependence of $\kappa_{\text{exp}}$ for the CoFe/[Cu/CoFe]$_{20}$ multilayer. We found that the CoFe/[Cu/CoFe]$_{20}$ multilayer exhibited a drastic $\kappa_{\text{exp}}$ change in the range from 0 mT to 25 mT, while $\kappa_{\text{exp}}$ remains almost constant at $\mu_0 H > 25$ mT. The field range exhibiting the large change in $\kappa_{\text{exp}}$ corresponds to the range where the transition from the minimum to the maximum resistance is observed in the MR curve [Fig. 2(a)]. This indicates that $\kappa_{\text{exp}}$ reaches its minimum and maximum values in the AP and P state, respectively, although the number of data points was limited. Although the observed MTR ratio (108%) was a smaller than the reported one (150%) in the previous study [19], the measured change in the thermal conductivity ($\Delta\kappa_{\text{exp}} = 37$ W m$^{-1}$ K$^{-1}$) was not only approximately 1.5 times larger than that reported in the previous study ($\Delta\kappa_{\text{exp}} = 25$ W m$^{-1}$ K$^{-1}$) but also the record-high value among solid-state thermal switching materials. In the previous report [19], the used Al thickness of 5nm may be insufficient for absorbing pump-laser energy, requiring the implementation of a bidirectional heat conduction model in TDTR analyses. In contrast, the present Al thickness (44 nm) was thick enough to assume the one-directional heat diffusion model. Despite different models used, both present and previous multilayers [19] exhibited giant MTR ratios, suggesting the robustness of the GMTR effect.

From the perspective of practical applications of thermal switching devices, achieving both low and high thermal conductivity states under zero external magnetic field (non-volatile bistability) is crucial. In the present multilayer, a clear hysteresis appears in both MR and magnetization curves [Figs. 2(a) and 2(b)], whereas no such hysteresis was observed in previous study [19]. The observed hysteresis probably originates from the increased crystalline magnetic anisotropy energy of CoFe due to the thicker CoFe layer thickness compared to the previous study. By utilizing this hysteretic behavior, we demonstrated the bistable control of the thermal conduction. As shown in Figs. 2(d) and 2(e), by adjusting $\mu_0 H$ to 10 mT with changing the sweeping direction of $H$, the TDTR curve and estimated $\Delta\kappa_{\text{exp}}$ value reproduced the both data at the AP and P states at one $\mu_0 H$ value. Although detailed mechanisms of non-volatile bistability are to be investigated, this result suggests that non-volatile bistability could be also obtained by stabilizing P and AP states at zero magnetic field, potentially through the exchange bias effect instead of IEC, exhibiting the feasibility of a nonvolatile thermal switching device based on the GMTR effect in the future.

## IV. DISCUSSION

Next, we analyzed the component of the electron thermal conductivity by simulating the *RA* (electrical resistance-area product) and MR ratio in the current-perpendicular-to-plane (CPP)-configuration using the generalized 2CSR model [Fig. 3(a)]. The 2CSR model [30-32] is the well-established theoretical approach that predicts the resistance change in the CPP-GMR structure by considering the individual series resistors for majority- and minority-spin channels. In the simulations, the CPP-GMR structure was modelled as Cu (1 nm)/[CoFe (5.1 nm)/Cu (2.0 nm)]$_{20}$/CoFe (5.1 nm)/Cu (1 nm), where the 1-nm-thick bottom and upper Cu layers considered as the electrodes. The simulation requires several parameters including the bulk resistivity ($\rho$), bulk spin-scattering asymmetry coefficient ($\beta$), spin diffusion length ($\lambda$), thickness of each layer ($t$), interfacial resistance area product ($r$), and interfacial spin-scattering asymmetry coefficient ($\gamma$) [33,34]. Reasonable parameters, such as $\rho$ of CoFe and Cu; $\beta$ of CoFe; $\lambda$ of CoFe and Cu, were adopted from previous studies [33,34], which were summarized in Table I. We used the $\rho$ and $\lambda$ reported for fcc-Cu in place of those for bcc-Cu as there is no reported values for bcc-Cu. It should be noted that these parameters in bulk region of Cu have negligibly weak effect on the simulated results because the thickness of Cu (2.0 nm) is much shorter than $\lambda$. However, it is necessary to obtain values of $r$ ($r_{\text{CoFe/Cu}}$) and $\gamma$ at the metastable CoFe/Cu interface ($\gamma_{\text{CoFe/Cu}}$) since the previous study has reported that the metastable bcc-Cu has a significantly different interfacial electronic band matching with CoFe compared to fcc-Cu [20]. In this study, we quantitively analyzed $r_{\text{CoFe/Cu}}$ and $\gamma_{\text{CoFe/Cu}}$ based on the experimental data of the [CoFe(3 nm)/bcc-Cu]$_N$ multilayer CPP-GMR devices [19]. By analyzing the measured $RA_{\text{AP}}$ for $N$ = 3, 5, and 7 [19] using the series circuit model consisting of the bulk and interfacial resistance area products, we determined $r_{\text{CoFe/Cu}}$ = 0.27 mΩ μm$^2$. We also attempted to obtain the $\gamma_{\text{CoFe/Cu}}$ by comparing the experimentally measured resistance change-area products ($\Delta RA$) for $N$ = 3, 5, and 7 [19] with the simulated $\Delta RA$ as a function of $\gamma_{\text{CoFe/Cu}}$ [Fig. 3(b)]. As a result, we found $\gamma_{\text{CoFe/Cu}}$ to be 0.76±0.03. By adopting these $r_{\text{CoFe/Cu}}$ and $\gamma_{\text{CoFe/Cu}}$ in the analysis based on the 2CSR model, we finally estimated $RA_{\text{AP}}$ = 47.2 mΩ μm$^2$, $RA_{\text{P}}$ = 29.4 mΩ μm$^2$, and obtained CPP-MR ratio of 60% for the present CoFe/[Cu/CoFe]$_{20}$ multilayer with the 5.1-nm-thick CoFe layers. The obtained MR ratio was much smaller than the experimentally observed MTR ratio, approximately half of the MTR ratio of 108%, whose tendency is consistent with the previous report [19]. Based on the WF law, the electron thermal conductivity of conventional metals is described as $\sigma LT$, where $\sigma$ ($= t_{\text{total}}/RA$) is the electrical conductivity, $L$ is the Lorenz number ($2.44 \times 10^{-8}$ W Ω K$^{-2}$), and $T$ is the absolute temperature. Assuming that the WF law can be applied for the present multilayer structure, the electron thermal conductivity values for the AP and P states ($\kappa_e^{\text{WF}}$) were estimated to be 22.6 and 36.2 W m$^{-1}$ K$^{-1}$ at $T$ = 297 K, respectively, using the simulated $RA_{\text{AP}}$ and $RA_{\text{P}}$ (see Table II). The calculated their difference $\Delta \kappa_e^{\text{WF}}$ of 13.6±0.5 W m$^{-1}$ K$^{-1}$ is only about 35% of the experimentally observed $\Delta \kappa_{\text{exp}}$ (37±10 W m$^{-1}$ K$^{-1}$). Thus, $\Delta \kappa_{\text{exp}}$ should include the additional spin-dependent contribution $\Delta \kappa_{\text{add}}$ (= $\Delta \kappa_{\text{exp}} - \Delta \kappa_e^{\text{WF}}$), which is calculated to be very large 24±10 W m$^{-1}$ K$^{-1}$. It should be mentioned that the $L$ value in ferromagnetic materials sometimes deviates from $2.44 \times 10^{-8}$ W Ω K$^{-2}$ due to several factors (e.g., spin-dependent scattering mediated by exchange interaction and high density of states near Fermi energy). However, the experimentally reported $L$ values in ferromagnetic 3$d$ transition metals at room temperature are slightly smaller than it [11,15]. By applying this scenario to our sample, intrinsic electron thermal conductivity in the present multilayer may be smaller than $\kappa_e^{\text{WF}}$, respectively. Therefore, it is expected that the deviation of $L$ increases the $\Delta \kappa_{\text{add}}$ component, but not decreases it. Another potential factor in correcting electron thermal conductivity is the Seebeck effect, expressed as $\sigma S^2 T$, where S is the Seebeck coefficient. However, assuming the $S$ values of our multilayer in the AP and P states are the same as those measured in the previous study [35], the correction on the electron thermal conductivity is estimated to be less than 1 W m$^{-1}$ K$^{-1}$ in both the AP and P states, which cannot explain the observed large $\Delta\kappa_{\text{add}}$.

Here, we explore the lattice component of thermal conductivity in CoFe/Cu multilayers. In actual CoFe layers, Co and Fe atoms are expected to be randomly alloyed, as annealing at 250°C is insufficient to form the ordered *B2* structure of CoFe [36]. Even though the layer thicknesses are on the nanometer scale, phonons are expected to experience significant alloy scattering. Furthermore, considering the Debye temperatures of the constituent elements [37], the quantum effect should have a minimal effect. Therefore, we employed NEMD simulations to evaluate the lattice thermal conductivity. As illustrated in Fig. 4(a), the CoFe/Cu multilayer was modeled by stacking a bcc-based CoFe/Cu slab along the *z* direction. The coordinates of Co and Fe atoms in each layer were randomly distributed while maintaining an atomic fraction of 1:1. Furthermore, their coordinates in different slabs were set to be uncorrelated [Fig. 4(b)]. The lattice constants of both CoFe and Cu slabs were set to 2.86 Å, as determined from the lattice spacing analysis of TEM images. The interatomic interactions among the Co, Fe, and Cu atoms were described by the embedded atom method potential with the parameterization of Zhou et al. [38]. In the NEMD simulation, a system with a length ($l$) and a cross-sectional area ($A$) was sandwiched by hot and cold thermostats with the length of $l/2$. The periodic boundary condition was applied to the cross-sectional direction. The temperatures of the thermostats were controlled by the Langevin thermostat. Denoting inflow and outflow of the energies in the thermostats at a time interval $\Delta t$ as $\Delta Q_H(\Delta t)$ and $\Delta Q_C(\Delta t)$, the heat flux in the system was given by the average; $q(t = k\Delta t) = \sum_{j}^{k}[\Delta Q_H(j\Delta t) - \Delta Q_C(j\Delta t)]/2Ak\Delta t$, where $k$ is the total time steps of the NEMD simulation. We first performed a 5 ns-long NEMD simulation to ensure the system reached a non-equilibrium steady state more smoothly. Subsequently, we conducted an additional 5 ns-long NEMD simulation to sample $q(t)$ and evaluated its mean value $\bar{q}$ was evaluated after a nonequilibrium steady state was established. An effective lattice thermal conductivity of the CoFe/Cu multilayer system was calculated by $\kappa_{\text{L}}^{\text{NEMD}} = \bar{q}l/\Delta T$, where $\Delta T$ denotes a temperature difference between the hot and cold thermostats. $\kappa_{\text{L}}^{\text{NEMD}}$ was finally evaluated through the ensemble average for five different atomic configurations. For the comparison with measurements, $l$ was set to approximately 143 nm. In addition, $A$ was set to 2.29×2.29 nm$^2$, which ensures the convergence of $\kappa_{\text{L}}^{\text{NEMD}}$. To fix the translational and rotational motions of the entire system, the CoFe/Cu slabs with frozen internal degrees of freedom were attached to both edges of the system. All NEMD simulations were performed by the LAMMPS package [39] with a time step of 0.5 fs. As shown in the Fig. 4(c), a steady state was achieved at 2.5 ns to 5.0 ns, $\kappa_{\text{L}}^{\text{NEMD}}$ was evaluated from the heat flux in the time range. Although small fluctuations appeared in the obtained temperature profilefor $\Delta T$ = 50 K [Fig. 4(d)], a nearly linear temperature profile was established, confirming the homogeneity of the CoFe/Cu multilayer. The resulting $\kappa_{\text{L}}^{\text{NEMD}}$ of the CoFe/Cu multilayer was 5.7 W m$^{-1}$ K$^{-1}$ at room temperature. The $\kappa_{\text{L}}^{\text{NEMD}}$ value remained independent of the magnitude of $\Delta T$ (see Supplementary Material [27]]. Note that the empirical interatomic potential used is not optimized for heat conduction in the present multilayer. This obtained value exhibits magnitude similar to previous superlattice calculations [40,41]. Since it is reasonable to assume the lattice thermal conductivity is independent of the magnetization configurations, we put the same $\kappa_{\text{L}}^{\text{NEMD}}$ for AP and P states and $\Delta\kappa_{\text{L}}^{\text{NEMD}} = 0$ in Table II.

Finally, we quantitatively compared the value of $\kappa_{\text{exp}}$, $\kappa_{\text{e}}^{\text{WF}}$, $\kappa_{\text{L}}^{\text{NEMD}}$, and explored the additional contribution of the thermal conductivity ($\kappa_{\text{add}} = \kappa_{\text{exp}} - \kappa_{\text{e}}^{\text{WF}} - \kappa_{\text{L}}^{\text{NEMD}}$) as summarized in Table II. The quantitative analysis of $\kappa_{\text{L}}^{\text{NEMD}}$ has led to an important finding: $\kappa_{\text{exp}}$ in the AP state is nearly equal to the summation of $\kappa_{\text{e}}^{\text{WF}}$ and $\kappa_{\text{L}}^{\text{NEMD}}$, i.e., $\kappa_{\text{add}} =$ 5±4 W m$^{-1}$ K$^{-1}$, whereas in the P state, it is significantly larger than their summation, i.e., $\kappa_{\text{add}} = 30\pm9$ W m$^{-1}$ K$^{-1}$. This result suggests that the contribution of additional heat carriers in AP state could be negligibly small, although our analysis still contains a certain level of the error bar. Notably, the present quantitative analysis shows that almost 42% of $\kappa_{\text{exp}}$ in the P state and 65% of $\Delta\kappa_{\text{exp}}$ originate from $\kappa_{\text{add}}$ that is neither the contributions of electron nor phonon calculated based on the WF law and the NEMD simulation, respectively. A possible additional carriers contribution to the GMTR effect is magnon, collective dynamics of localized magnetic moment in magnetic materials. The magnetization-configuration-dependent magnon transport, known as the magnon valve effect, has been demonstrated in magnetic multilayer structures, which highlights the importance of magnon–magnon interactions between adjacent magnetic layers [42,43]. Although the magnon contribution in the thermal conduction is usually discussed at the low temperature [44-46], recent theoretical calculation has shown that the magnon thermal conductivity also depends on the magnetization configuration even at room temperature [47]; in the trilayer consisting of two magnetic insulator layers sandwiching a Cu interlayer, the large magnon-driven MTR ratios of up to 40% have been predicted. However, even if magnons contribute to $\kappa_{\text{exp}}$ in the P state, their thermal conductivity in typical ferromagnetic metals is around ~10 W m$^{-1}$ K$^{-1}$ at room temperature [28], which is much smaller than the estimated $\kappa_{\text{add}}$. Therefore, the conventional sole magnon contribution is unlikely to be a major factor in our observations. Thus, we should also consider the possibility of magnon-electron and magnon-phonon interactions [48-50]. An example of a phenomenon to which such interactions contribute is the magnon-drag effect in the thermoelectric conversion [48,49]. However, the reported modulation of thermopower by the magnon origin was very small in ferromagnetic metals, such as CoFe- and NiFe-based system, at room temperature [35,49]. If the magnon-drag-induced change in the thermal conductivity in our CoFe/Cu multilayer is comparable to that in the thermopower, its contribution is negligible. Regarding phonons, there is a possibility that they are influenced by the external magnetic field through spin-lattice interaction [51,52]. However, changes in phonon thermal conductivity have been observed only at the low temperature and under the strong magnetic field, which are completely different from the conditions in this study. Moreover, these changes were on the order of 0.1 W m$^{-1}$ K$^{-1}$, making their impact relatively small. We cannot entirely rule out the possibility of that some of these magnon-related interactions may still be significant, as we have not explicitly evaluated their effects in our sample. To clarify the microscopic mechanism of giant MTR ratio and $\kappa_{\text{exp}}$, further experiments and calculations from a series of multifaceted perspectives are necessary.

## V. CONCLUSIONS

In summary, we investigated the GMTR effect in the CoFe/Cu magnetic multilayer with 5.1 nm-thick CoFe layers. We achieved the giant thermal conductivity change of 37 W m$^{-1}$ K$^{-1}$, not only surpassing the previous study [19] but also marking the highest value in the solid-state thermal switching materials. The MTR ratio of 108% is found to be greater than the MR ratio for the CPP configuration, reproducing the tendency of previous study for 3 nm-thick CoFe. By comparing the value of $\kappa_{\text{exp}}$, $\kappa_{\text{e}}^{\text{WF}}$, and $\kappa_{\text{L}}^{\text{NEMD}}$, we identified a significant contribution from additional components of the thermal conductivity to the GMTR effect although this contribution has not been uncovered yet in this study.

Interestingly, the quantitative analysis of the lattice thermal conductivity suggests that the contribution of this additional heat carriers is remarkable in P state but negligibly small in AP state. This could provide an important clue to elucidate the origin of the unconventional thermal transport in the magnetic multilayers. Furthermore, this study demonstrated the bistability of low and high thermal conductivity states at the same magnetic field due to the hysteretic behavior of the magnetization. These findings highlight the potential of active thermal management technologies using the GMTR effect for electronic devices.

**ACKNOWLEDGEMENTS**

The authors thank G. E. W. Bauer, P. Tang, Y. Yamashita, Y. Miura, R. Iguchi, and W. Zhou for valuation discussion and S. Kuramochi and N. Kojima for the technical support. This work was supported by JST-ERATO "Magnetic Thermal Management Materials Project" (No. JPMJER2201); JST-CREST "Creation of Innovative Core Technology for Nano-enabled Thermal Management" (No. JPMJCR17I1); JST-FOREST (JPMJFR222G), JSPS KAKENHI (Nos. 22K20495 and 22H04965); and NIMS Joint Research Hub Program. T.H. acknowledged support from the Thermal and Electric Energy Technology Foundation. A part of the present calculations was performed on the TSUBAME4.0 supercomputer at Institute of Science Tokyo.

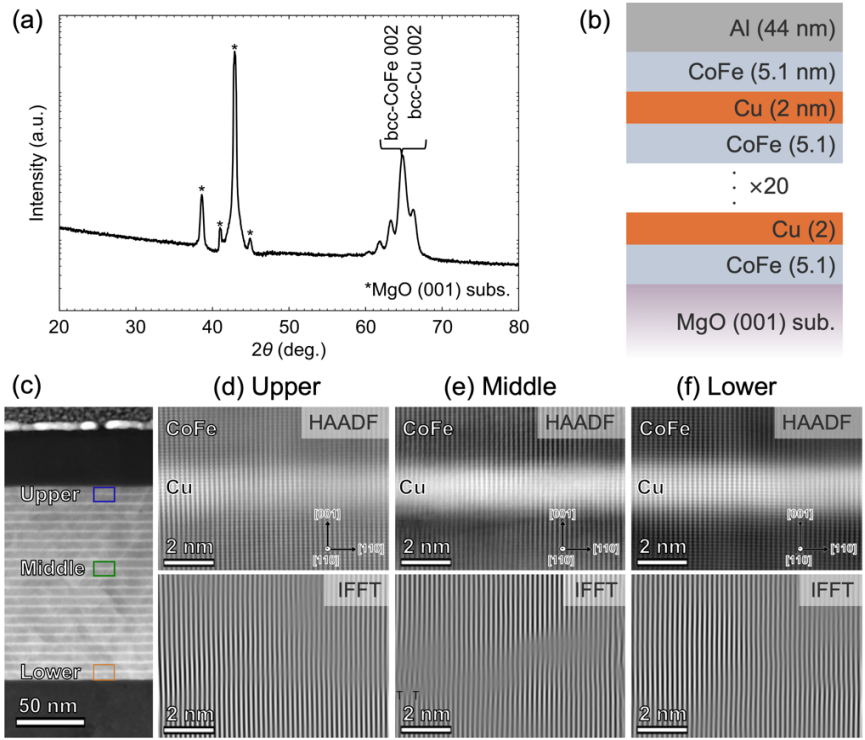

FIG.1. (a) Out-of-plane x-ray diffraction curve of the multilayer film. (b) Schematic structure of the CoFe/[Cu/CoFe]$_{20}$ multilayer film deposited on the MgO (001) single crystalline substrate. (c) Low-magnification high-angle annular dark field scanning transmission electron microscopy (HAADF-STEM) image of the multilayer film. (d)-(f) High-magnification HAADF-STEM images and inverse fast Fourier transform (IFFT) images reconstructed from 110 peaks of (d) Upper, (e) Middle, and (f) Lower parts indicated by the rectangular regions in (c), respectively.

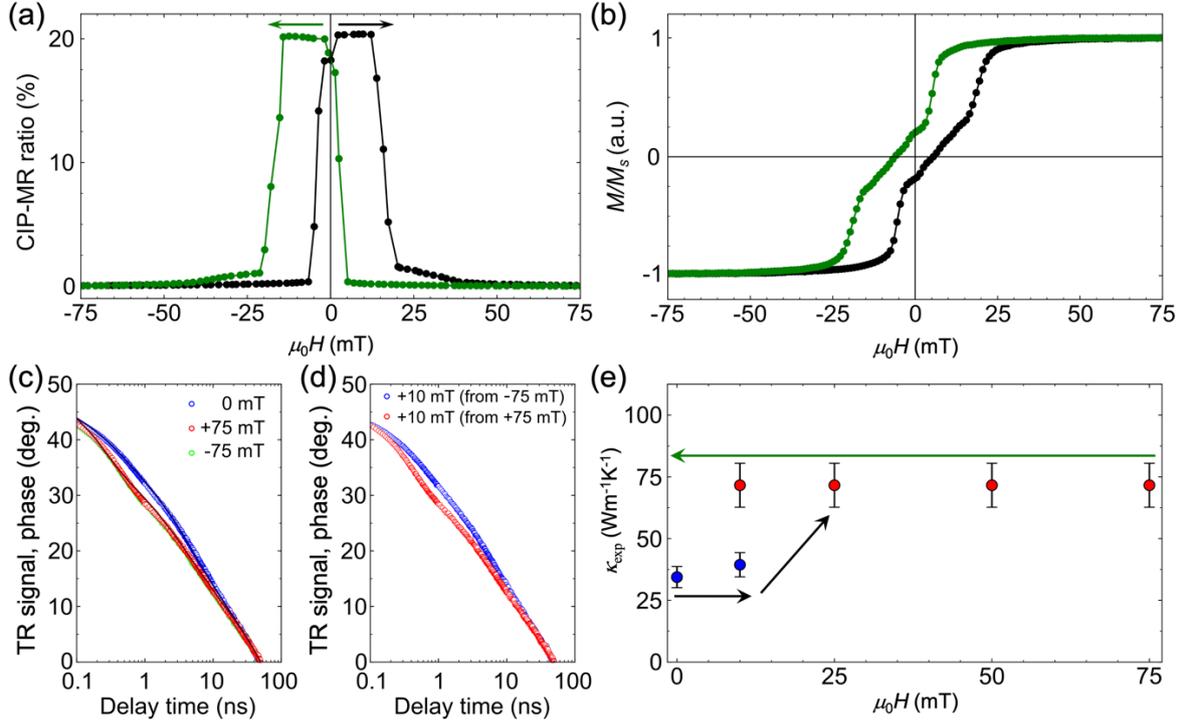

FIG.2. (a) Magnetic field $H$ dependence of current-in-plane magnetoresistance (CIP-MR) and (b) normalized magnetization for the CoFe/[Cu/CoFe]$_{20}$ multilayer film with antiferromagnetic coupling via Cu spacer. (c) Temporal response of thermoreflectance (TR) signals, i.e., TDTR signals, for the CoFe/[Cu/CoFe]$_{20}$ multilayer film with applying the magnetic field $\mu_0 H$ of 0 and $\pm 75$ mT. The black solid curves represent the best fits for each measurement. (d) TDTR signals for the CoFe/[Cu/CoFe]$_{20}$ multilayer film at $\mu_0 H$ of 10 mT applied with adjusting the field sweep direction. (e) $H$ dependence of $\kappa_{\text{exp}}$ of the CoFe/[Cu/CoFe]$_{20}$ multilayer film. Black and green arrows in (a) and (e) indicate the sweep direction of the magnetic field

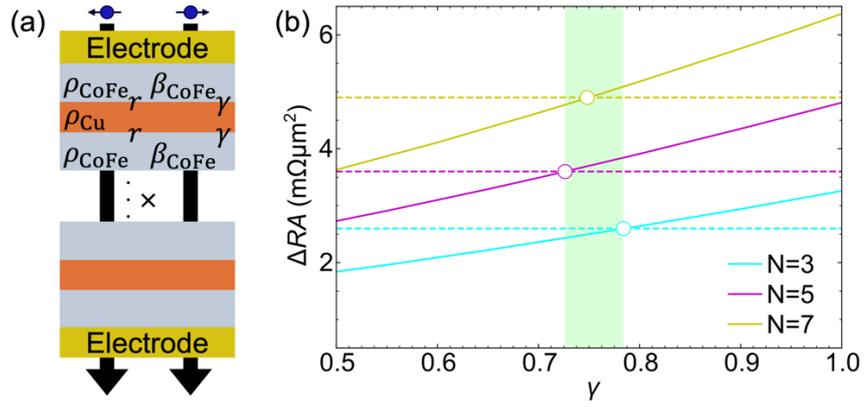

FIG.3. (a) Schematic of two-current series-resistor (2CSR) model. (b) Simulated spin interfacial asymmetry $\gamma$ dependence of resistance change-area product $\Delta RA$ (solid curves). Dashed lines show experimentally measured $\Delta RA$ of the CoFe/[Cu/CoFe]$_N$ multilayer with repeating number $N$ of 3, 5, and 7 [15]. Open circles show the $\gamma$ values obtained from the agreement between experimental and simulated $\Delta RA$.

TABLE I. Used parameter for the generalized two-current series-resistor (2CSR) model

|      | $\rho$ [μΩ cm] | $\beta$   | $\lambda$ [nm] | $t$ [nm] | $r$ [mΩ μm$^2$] | $\gamma$    |
|------|----------------|-----------|----------------|----------|------------------|-------------|
| CoFe | 19.1 [22]      | 0.62 [23] | 15 [22]        | 5.1      | 0.27             | 0.76±0.03   |
| Cu   | 7.0 [22]       | 0         | 100 [22]       | 2.0      | 0.27             | 0.76±0.03   |

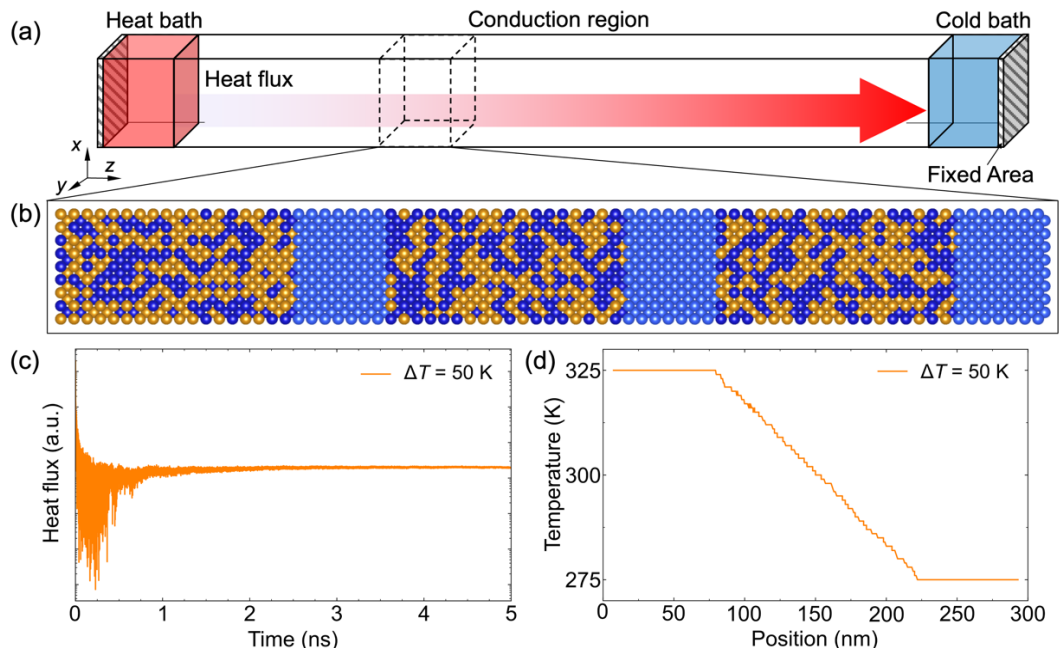

FIG.4. (a) Schematic of the model for the nonequilibrium molecular dynamics (NEMD) simulations. (b) Stacking structure for NEMD. The atomic position is determined so as to be different from the adjacent CoFe layers. (c) Time dependence of the heat flux for temperature difference, $\varDelta T$ of 50 K. (d) The obtained temperature profile for $\varDelta T$ of 50 K and the liner temperature approximation used to calculate the temperature gradient.

TABLE II. Summary of the experimentally observed thermal conductivity $\kappa_{\text{exp}}$, the theoretically analyzed electron thermal conductivity $\kappa_{\text{e}}^{\text{WF}}$ and effective lattice thermal conductivity $\kappa_{\text{L}}^{\text{NEMD}}$ and the evaluated additional component of thermal conductivity $\kappa_{\text{add}}$ in AP and P states.

|  | $\kappa_{\text{exp}}$ [W m$^{-1}$ K$^{-1}$] | $\kappa_{\text{e}}^{\text{WF}}$ [W m$^{-1}$ K$^{-1}$] | $\kappa_{\text{L}}^{\text{NEMD}}$ [W m$^{-1}$ K$^{-1}$] | $\kappa_{\text{add}}$ [W m$^{-1}$ K$^{-1}$] |
|---|---|---|---|---|
| AP state | 34±4 | 22.6 | 5.7±2.0 | 5±4 |
| P state | 72±9 | 36.2±0.5 | 5.7±2.0 | 30±9 |
| Δκ | 37±10 | 13.6±0.5 | 0 | 24±10 |